\documentclass[lettersize,journal]{IEEEtran}
\usepackage{amsmath,amsfonts}
\usepackage{algorithm}
\usepackage{algorithmic}
\usepackage{array}
\usepackage[caption=false,font=normalsize,labelfont=sf,textfont=sf]{subfig}
\usepackage{textcomp}
\usepackage{stfloats}
\usepackage{url}
\usepackage{verbatim}
\usepackage{graphicx}
\usepackage{xcolor}
\def\BibTeX{{\rm B\kern-.05em{\sc i\kern-.025em b}\kern-.08em
    T\kern-.1667em\lower.7ex\hbox{E}\kern-.125emX}}
\usepackage{balance}
\begin{document}
\title{Generative Communications: Overview, Technologies, and Trends}
\author{Wenjun Zhang, \emph{Fellow, IEEE}, Zhiyong Chen, \emph{Senior Member, IEEE}, Tong Wu, Guo Lu, \emph{Member, IEEE}, \\Li Song, \emph{Senior Member, IEEE},  Feng Yang, and Meixia Tao, \emph{Fellow, IEEE}
\thanks{The authors are with the Cooperative Medianet Innovation Center and the School of Information Science and Electronic Engineering, Shanghai Jiao Tong University, Shanghai 200240, China. (e-mail:\{zhangwenjun, zhiyongchen, wu\_tong, luguo2014, song\_li, yangfeng, mxtao\}@sjtu.edu.cn). (Corresponding author: Zhiyong Chen, Tong Wu.)}}


\maketitle

\begin{abstract}
The groundbreaking development of generative artificial intelligence (AI) is rapidly boosting the ability to generate content such as images and videos, reshaping communication paradigms. This article introduces generative communications (GenCom), a novel paradigm for 6G networks in which large AI models (LAMs) drive semantic understanding, reasoning, and content generation, embedding these into the communication process. Unlike traditional systems that strictly pursue accurate bit transmission, GenCom enables transmitters to convey only minimal yet sufficient information, while receivers leverage shared generative priors and knowledge bases to synthesize the intended output. Communication is thus redefined as controlled generation rather than data reproduction. We formalize the concept of GenCom, clarify its AI-native and generation-driven properties, and present its core mechanisms. A two-layer GenCom architecture supported by key enabling technologies is proposed, and analysis of four representative application scenarios demonstrates that GenCom offers ultra-efficient transmission, semantic-level robustness, and new network functions. Finally, we outline future research directions, including foundational theory and real-time processing, highlighting a promising pathway toward 6G networks.
\end{abstract}


\section{Introduction}
As we advance toward the sixth-generation (6G) mobile communication era, the focus of network design is shifting from device-centric interconnection to intelligent interaction among humans, machines, and environments. 6G is envisioned not only to provide extreme connectivity and ultra-high performance but also to be AI-native, endowing the network with capabilities for semantic understanding, autonomous decision-making, and real-time self-adaptation \cite{Jiang2024Large}. Consequently, communication is expected to evolve beyond the pursuit of efficient bit transmission, moving toward knowledge-driven and task-oriented information exchange.

However, the current communication system, fundamentally based on Shannon's theory, defines the communication process as the accurate transmission of information from transmitter to receiver, with its core design objective being bit transmission efficiency. This framework achieved immense success in traditional communication services. However, its limitations become apparent in intelligent scenarios with growing cognitive demands in the 6G era. Firstly, this paradigm lacks inherent semantic understanding capability, as it focuses exclusively on low-level metrics like bit error rate (BER) and throughput, thereby inherently overlooking the semantic meaning and task relevance of the information. Secondly, communication and intelligence are typically decoupled in existing networks, preventing systems from actively comprehending, predicting, or generating content during transmission. This decoupling leads to inefficient resource utilization, as the transmission of numerous bits is often irrelevant to the final task objective, resulting in semantic redundancy and wasted bandwidth.

\begin{figure}
    \centering
    \includegraphics[width=1\linewidth]{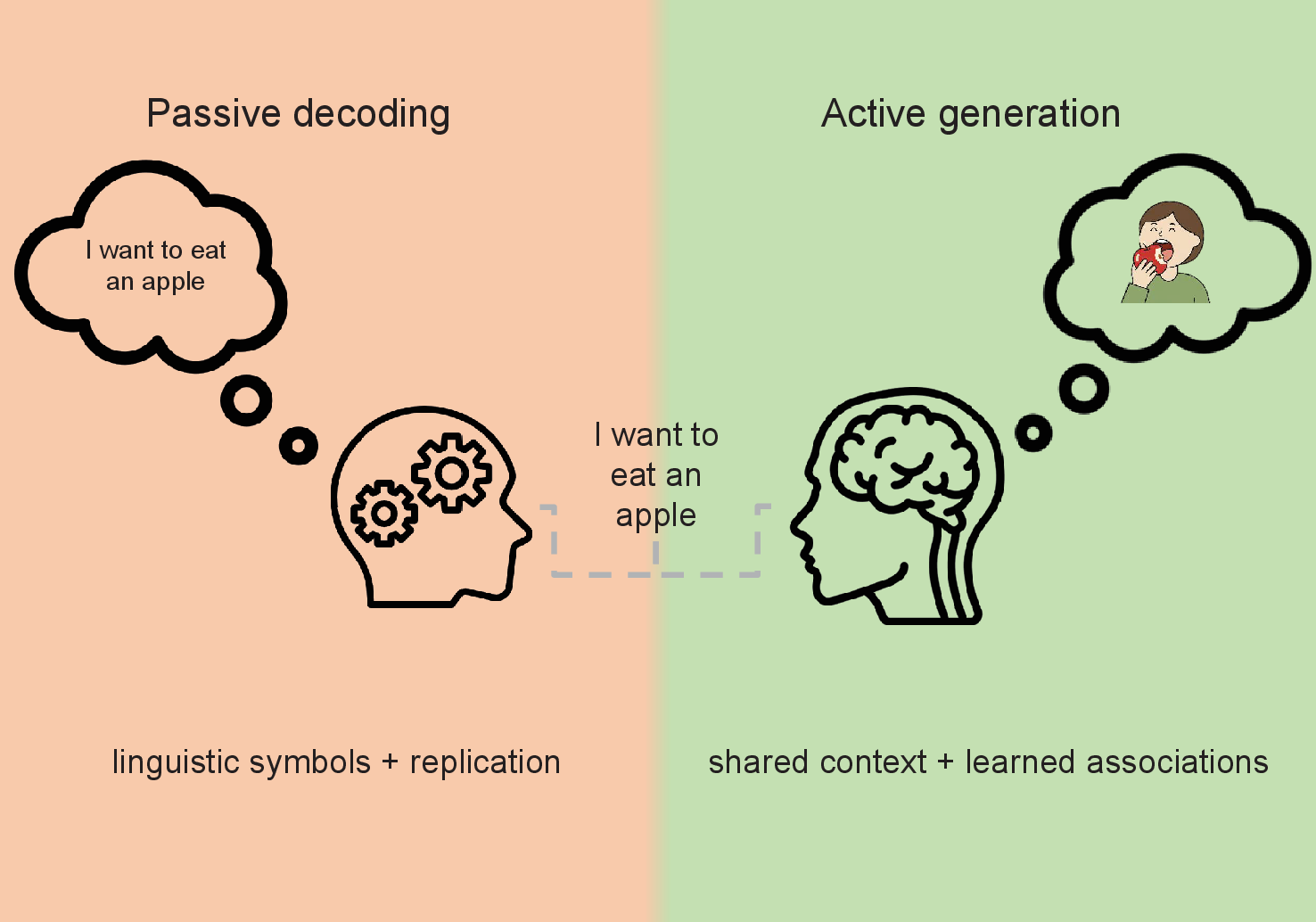}
    \caption{An example of human understanding in the brain.}
    \label{fig:cognitive paradig}
\end{figure}

From a cognitive perspective, human understanding itself is inherently \textbf{generative} \cite{clark2013whatever}. When perceiving or hearing a concept, the brain does not simply decode linguistic symbols but actively reconstructs a mental representation of the object based on prior knowledge and experience, as shown in Fig. \ref{fig:cognitive paradig}. This process is guided by internal cognitive schemas—structured knowledge templates that enable meaning reconstruction rather than mere symbol replication. In essence, humans achieve understanding not by receiving all the data, but by generatively recreating meaning through shared context and learned associations. This cognitive mechanism provides an important inspiration for next-generation communication systems \cite{ZHANG2025Generative}: the ability to achieve effective understanding and task completion with minimal data exchange, by leveraging shared semantic priors and generative reasoning.

Motivated by this cognitive paradigm, future communication systems should evolve toward machine-level generative understanding, where meaning is co-created rather than transmitted. This vision is being materialized through the rapid advancement of artificial intelligence (AI), particularly generative AI (GenAI), which offers a powerful engine to transform communication from passive transmission to active generation. The recent breakthroughs in GenAI have fundamentally redefined the boundaries of machine intelligence, enabling systems not only to analyze and predict but also to create new content with remarkable realism and coherence. With large language models (LLMs) and diffusion models (DMs), GenAI has demonstrated extraordinary capabilities in multi-modal generation—spanning text, image, audio, and even complex symbolic reasoning. 
\begin{figure*}
    \centering
    \includegraphics[width=1\linewidth]{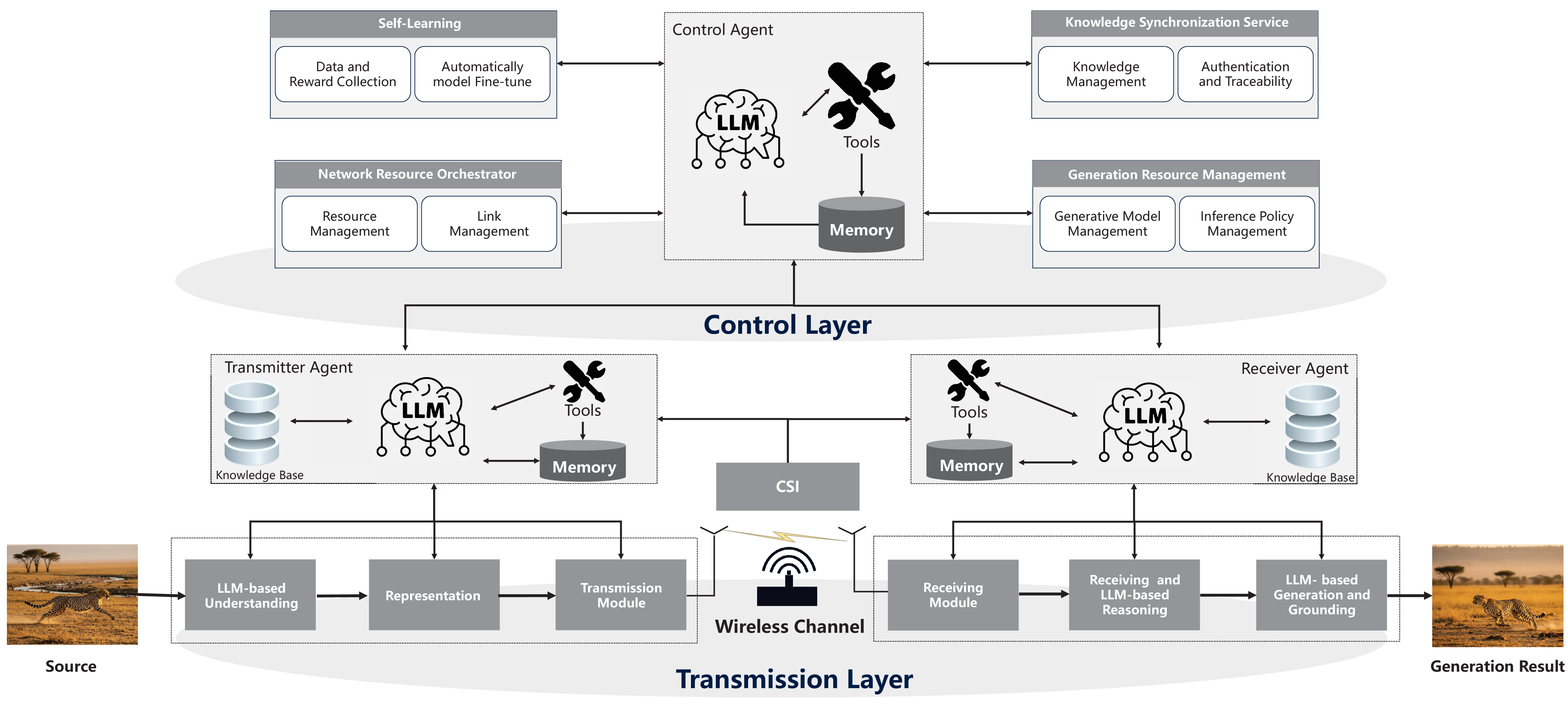}
    \caption{A structure of generative communications for 6G Networks.}
    \label{structure of GenCom}
\end{figure*}

Therefore, this paper introduces a new communications paradigm, termed as \textbf{generative communications (GenCom), which integrates understanding, reasoning, and generation directly into the communication process}. GenCom aims to transform communication from passive information transmission to active meaning generation, enabling networks to comprehend intent and autonomously produce task-relevant information rather than merely conveying raw data.

GenCom opens transformative opportunities across multiple layers of 6G networks. At the physical layer, it supports intent-oriented representations and cross-modal content generation, fundamentally redefining how information is transmitted and reconstructed. Communication systems are no longer constrained by bit-level accuracy or exact source reproduction. Instead, by leveraging shared generative priors and knowledge bases at both the transmitter and receiver, the receiver can perform controlled, creative generation that fulfills the intended communicative goal. At the network layer, GenCom integrates intelligent agents into resource scheduling and orchestration, establishing AI-native mechanisms for dynamic, self-learning, and autonomous management across network domains.

Although GenCom presents broad application prospects for future 6G networks, its theoretical framework and key technological pathways are still in the early stages of exploration. At present, GenCom faces numerous challenges in fundamental theory, system architecture, and core technologies, leading to the absence of a comprehensive technical overview and systematic analysis of future trends. To fill this gap, this paper aims to provide a thorough overview of GenCom. The next section presents GenCom in terms of its definition, core mechanisms, and advantages. We then propose the architecture and key technologies for GenCom, followed by four typical application scenarios that illustrate its potential. Moreover, we outline promising directions for future research. Finally, we conclude the article.

\section{What is generative communications?}
\subsection{Definition}
Inspired by the cognitive property, we define generative communications as a communication paradigm that embeds understanding, reasoning, and generation directly into the communication process, enabling transmitter and receiver to collaboratively produce task-relevant outcomes, as shown in Fig. \ref{structure of GenCom}. In GenCom, the transmitter does not convey a full description of the content; instead, it sends minimal yet sufficient cues that, when combined with the receiver’s shared generative priors and reasoning capabilities, allow the receiver to generate the desired result. Communication thus transforms from reproducing transmitted data to the controlled generation of desired outputs.

GenCom is fundamentally characterized by two core properties as follows:
\begin{itemize}
\item{AI-native: GenCom incorporates AI capabilities—such as semantic understanding, logical reasoning, and multimodal generation—into the communication process itself. These functions are not peripheral add-ons but essential components governing how information is represented, transmitted, and reconstructed. The receiver evolves from a passive endpoint into an intelligent generator equipped with pretrained models (e.g., LLMs, DMs).}

\item{Generation-driven: In GenCom, the transmitted signal may be highly compressed and may encode only intents, constraints, or abstract latent cues. The essence of communication lies in how effectively the receiver can transform these cues into high-quality, task-aligned outputs using its generative priors. Accordingly, the value of communication is determined not by symbol fidelity but by the quality and correctness of the generated result.}
\end{itemize}

\subsection{The Core Mechanisms}
The core mechanism of GenCom lies in the deep fusion of three elements: transmitted information, generative model priors, and knowledge bases. Their integration forms a controlled generation pipeline: the transmitted information provides the conditioning signal, the generative model priors supply the structural and semantic generative capability, and the knowledge bases ensure factual grounding. Through this interplay, the receiver is no longer constrained by the transmitted data itself but can reconstruct, infer, and generate high-quality outputs via shared priors and external knowledge resources.

\subsubsection{Transmitted Information}

\begin{table*}[t]
\caption{Representative signaling formats for transmitted units in GenCom and their qualitative tradeoffs.}
\label{tab:signaling_formats_gencom}
\centering
\footnotesize
\renewcommand{\arraystretch}{1.22}
\setlength{\tabcolsep}{4.2pt}
\begin{tabular*}{\textwidth}{@{\extracolsep{\fill}}>{\raggedright\arraybackslash}p{2.35cm}>{\raggedright\arraybackslash}p{3.25cm}>{\centering\arraybackslash}p{1.55cm}>{\centering\arraybackslash}p{1.55cm}>{\centering\arraybackslash}p{1.75cm}>{\raggedright\arraybackslash}p{4.05cm}@{}}
\hline
\textbf{Signaling format} & \textbf{Examples} & \textbf{Payload} & \textbf{Robustness} & \textbf{Control} & \textbf{Typical use} \\
\hline
Textual semantic & Prompt, caption, task instruction & Low & High & Low--Medium & Semantic-consistent generation under tight bandwidth \\
\hline
Structured signal & Segmentation map, subject cue, layout, mask & Low--Medium & Medium--High & High & Task-oriented generation with explicit control \\
\hline
Latent representation & Latent code, compressed embedding, steering vector & Medium & Medium & Medium--High & Model-aligned generation balancing compactness and control \\
\hline
Visual observation & Downsampled image, partial object, compressed frame & Medium--High & Medium & High & High-fidelity generation with stronger visual consistency \\
\hline
\end{tabular*}
\end{table*}
The data sent through the physical channel is no longer a full representation of the original message. Instead, it is minimized to contain only the essential cues—such as semantic descriptors, constraints, or residual updates—that condition and guide the generation process at the receiver. Its role is to act as a triggering or control signal, enabling highly efficient communication. For example, rather than transmitting all pixels of a high-resolution image, the sender may send only a short prompt such as “Emperor penguin on a glacier” along with a style code.

The transmitted unit in GenCom should not be interpreted as a fixed data type. Its concrete form depends on the task objective, the required fidelity, and the control interface of the generative model. It may therefore appear as a text prompt, a latent code, a segmentation map, a subject cue, or a compact visual observation such as a downsampled image. What is transmitted describes the generative-level content to be conveyed, while how it is delivered can still rely on conventional coding or generation-aware schemes such as JSCGC.Table \ref{tab:signaling_formats_gencom} summarizes representative signaling formats, possible payload size, robustness, and controllability.

\subsubsection{Generative Model Priors}
The generative model priors represent the shared internal structures, latent spaces, and parametric knowledge of the underlying intelligent models (e.g., LLMs or DMs) pre-deployed at both ends. These priors are not transmitted during communication; instead, they provide the generative backbone that allows the receiver to expand sparse input cues into rich, coherent content. By leveraging these shared priors, the receiver can perform semantic inference, controlled synthesis, and extrapolation far beyond what is contained in the minimal transmitted data.

In practice, generative priors may include pre-trained model weights, shared latent structures, domain knowledge bases, and communication context such as historical interactions or task states. The model weights encode data distribution and structural priors, the knowledge bases provide factual grounding, and the communication context helps adapt generation to the current scenario. These priors are not transmitted during communication; instead, they provide the generative backbone that allows the receiver to expand sparse input cues into rich and coherent content.

\subsubsection{Knowledge Bases}
Knowledge bases, whether external or locally maintained, provide the factual and contextual grounding needed to ensure accuracy and reliability. They supply domain knowledge, contextual information, and verified facts that calibrate and validate the generative output. When combined with the model priors and transmitted cues, the knowledge bases reduce hallucination and enforce consistency with real-world constraints. They also ensure that the generated content is trustworthy and contextually appropriate.

When the transmitter-side knowledge base conflicts with the receiver's local knowledge base, the inconsistency should be handled through the knowledge synchronization service in the control layer. The system can detect conflicts through knowledge-base versions, knowledge-delta updates, and provenance records, and then trigger synchronization or updating procedures. If immediate consistency cannot be guaranteed, a conservative strategy should be adopted, such as prioritizing the authorized knowledge source, delaying generation, or restricting the output to avoid unsafe or ungrounded results.

In summary, GenCom achieves its functionality through the coordinated fusion of minimal transmitted cues, shared generative priors, and factual knowledge. This triad enables the receiver to construct high-quality content that far exceeds the information explicitly sent, shifting communication from data reproduction to controlled generation \cite{Chaccour2025Less}.

\subsection{Advantages of GenCom}
GenCom brings transformative advantages fundamentally distinct from those of traditional communication systems. By shifting the objective from reproducing transmitted data to generating task-aligned outputs, GenCom unlocks new efficiency, scalability, and intelligence properties across the communications system.
\subsubsection{Ultra-efficient Transmission} GenCom radically reduces communication overhead by transmitting only minimal semantic cues rather than full data streams. High-dimensional content is reconstructed at the receiver using shared generative priors, enabling substantial bandwidth savings.
\subsubsection{Semantic-level Robustness}
Because reconstruction relies on meaning rather than exact symbols, GenCom is inherently robust to channel distortions. As long as the transmitted cues preserve core semantics, the receiver can regenerate coherent and high-quality outputs despite noise or partial loss.
\subsubsection{Enabling New Network Functions}
By integrating reasoning, generation, and knowledge grounding into the communications systems, GenCom supports new types of network functions previously unattainable, e.g., on-device content synthesis, autonomous decision support, dynamic task execution, and context-aware multi-agent collaboration. Communications become an active computational capability rather than a passive transport medium.

The above definition and core mechanism naturally motivate the two-layer architecture in Section III, where the transmission layer realizes controlled generation at the signal level, and the control layer maintains knowledge, model, and resource coordination at the system level.

\section{Architecture and Key Technologies}
\subsection{Architecture}
As illustrated in Fig. \ref{structure of GenCom}, we present a network architecture for GenCom in 6G networks. The architecture consists of two collaborative layers: the Transmission Layer, which performs semantic understanding and compact representation at the transmitter and inference-driven generative reconstruction at the receiver; and the Control Layer, which provides global coordination of knowledge, model, and resource management to ensure consistent and efficient network operation.

\textbf{Transmission Layer}: The transmission layer embodies the core GenCom pipeline between distributed terminals. At the transmitter, the source signal is first processed by an LLM-based understanding module, which interprets its semantic content and extracts task-relevant meaning. Based on this understanding, the representation module generates a compact trigger or latent representation that captures only the essential information required for downstream reconstruction. This minimal representation is delivered through the transmission module over the wireless channel.

At the receiver, the transmitted representation is decoded and fed into an LLM-based reasoning module, which infers the underlying intent or latent structure. Leveraging the local generative model, knowledge base, and memory, the receiver then performs generation and grounding to synthesize the final output that aligns with both the transmitted cues and world knowledge. This layer realizes the principle of generative communication, in which the burden of reconstruction shifts from the transmitted data itself to shared semantic priors and generative capabilities at both terminals.

\textbf{Control Layer}: The control layer provides global orchestration, ensuring consistency, adaptability, and optimized utilization of generative resources across the network. At its core lies the control agent, equipped with LLM, memory and tool interfaces, which coordinates multiple control services.

The knowledge synchronization service maintains coherence and version control of distributed knowledge bases, while also providing authentication and traceability for shared semantic priors. The generation resource management module oversees the deployment, updating, and scheduling of generative models, e.g., LLMs or DMs, and governs model selection and inference policies based on computational constraints. In parallel, the network resource orchestrator manages communication resources such as spectrum, links, and radio configurations, adapting them to the unique characteristics of generative workloads. Furthermore, the self-learning component continuously collects operational data and reward signals to refine model parameters and control policies, enabling the network to improve autonomously over time.

More concretely, the knowledge synchronization service exchanges knowledge-base versions, knowledge-delta updates, and model or adapter versions. These control items are typically slow-timescale or event-driven, with low frequency but potentially larger overhead per update. Generation resource management exchanges generator selection, inference mode, generation fidelity, controllability level, and compute budget, leading to light-to-moderate overhead at an intermediate timescale. Network resource orchestration handles CSI, spectrum, bandwidth, links, radio configurations, and compute scheduling through lightweight but more frequent updates. Self-learning collects task success, reward, and latency statistics in a periodic or event-driven manner, and the associated overhead is moderate and compressible.

Trust, safety, and security should also be treated as intrinsic functions of the control layer. Trust requires version consistency, provenance marks, and traceability metadata for shared knowledge resources, models, and generated outputs. Safety requires supervision and fallback mechanisms so that the system can verify whether the current generation strategy satisfies task and network constraints, and then switch to lighter models, stricter control modes, or bypass generation when needed. Security requires content-level supervision against malicious manipulation, privacy leakage, and violations of legal, ethical, or service-level constraints, together with feedback actions such as output rejection, regeneration under stricter constraints, or termination of the generation process.

The Control layer is also responsible for resource allocation in multiuser GenCom. In the scenario, resource allocation is no longer a radio-only problem. The system should jointly allocate spectrum, bandwidth, power, and time-frequency slots together with inference capacity and compute budget according to user intent, channel conditions, latency requirements, device capability, and target generation quality. Correspondingly, fairness should be evaluated by user-level service outcomes such as task utility, generation quality, latency satisfaction, and task success rate, rather than by throughput alone.

A brief and representative example is an XR hotspot where the number of active users suddenly increases. The network resource orchestrator first detects the surge in radio and compute load and reallocates bandwidth, radio resources, and compute budgets across users. The generation resource management module then adapts the generation strategy, for example by switching some users to lighter models or lower-fidelity generation modes to maintain service continuity. Meanwhile, the knowledge synchronization service keeps edge nodes and terminals consistent in their knowledge-base and model versions for the shared XR scene priors, and the self-learning module later uses task success, latency, and user-experience feedback to refine future control policies.

In summary, the two layers interact closely: terminals in the transmission layer rely on the control layer for model provisioning, knowledge consistency, and adaptive resource allocation, while the control layer gathers feedback and operational statistics from terminal activities to optimize system-wide performance. This cross-layer interaction establishes a unified, model-driven communication paradigm where meaning reconstruction, resource allocation, and knowledge maintenance are jointly optimized.
\begin{figure*}
    \centering
    \includegraphics[width=1\linewidth]{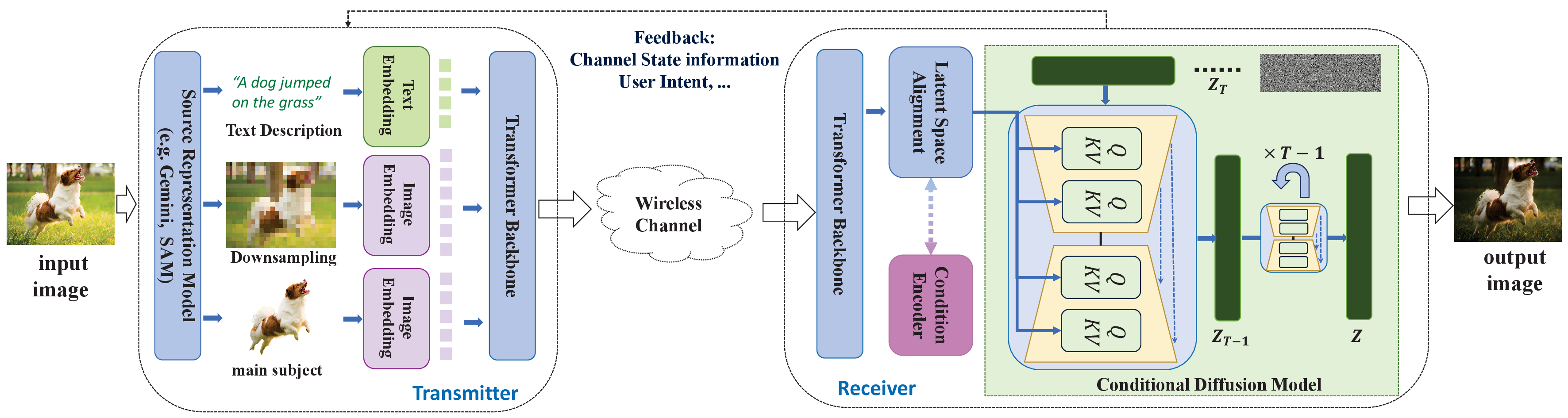}
    \caption{A framework of the proposed JSCGC. The transmitter first extracts multimodal control information, including the text description, downsampled image, and main subject. These representations are jointly encoded and transmitted through the wireless channel. At the receiver, the recovered representation is mapped into an aligned latent space, injected into the conditional encoder, and then guide the conditional diffusion model to generate images.}
    \label{JSCGC}
\end{figure*}

\subsection{Key Technologies}
The realization of GenCom relies on a set of foundational technologies that collectively enable the shift from data reproduction to controlled, knowledge-guided generation. The main enabling technologies are summarized as follows.

\textbf{Joint Source-Channel-Generative Coding (JSCGC):} JSCGC \cite{JSCGC, CDDM} unifies source representation, channel robustness, and generative conditioning into a single coding framework. Unlike traditional coding schemes that aim to minimize bit errors or preserve semantic fidelity, JSCGC directly optimizes the transmitted signal for generation compatibility. The encoded representation is crafted to align with the latent space and reasoning pathways of the receiver’s generative model, ensuring that the model can reliably interpret the information and synthesize the intended output—even under wireless channels. This shifts communication from passive reconstruction to active generative enablement. Potential realizations include latent-aligned encoding, diffusion-based condition coding, prompt–channel mapping, and model-aware noisy-channel co-training. 

For clarity, JSCGC should be understood as a generation-oriented coding framework rather than a conventional coding pipeline followed by a separate generator. Its objective is to jointly design source coding, channel coding, and the generation interface so that, under limited wireless resources such as power and bandwidth, the transmitted signal remains effective for controlling the receiver-side generative model. The transmitter first extracts task-relevant control information and jointly encodes it into generation-oriented channel symbols. After wireless transmission, the receiver recovers the latent representation, aligns it with the conditional embedding space used by the generative model, and then injects the aligned representation through an appropriate conditioning interface. We provide a brief high-level pipeline of JSCGC in Algorithm \ref{alg:jscgc} for clarity.

To illustrate the practical feasibility of JSCGC, we present a concrete implementation framework based on conditional diffusion models, as depicted in Fig .\ref{JSCGC}. In this architecture, the transmitter leverages foundation models (e.g., Qwen-VL and SAM ) to extract a multi-modal representation comprising the text description, downsampled image, and the main subject. These representations are jointly encoded and transmitted through the wireless channel via embedding and Transformer backbones. The coding scheme takes the generation process at the receiver into consideration, aiming to serve for the cross attention module in the Conditional Diffusion Model for guiding it to generate the desired output after the Transformer backbone and space alignment.
\begin{algorithm}[t]
\caption{High-level pipeline of JSCGC}
\label{alg:jscgc}
\begin{algorithmic}[1]
\REQUIRE Source content $x$, task requirement $r$, wireless channel $h$
\ENSURE Generated output $\hat{y}$
\STATE Extract task-relevant control information $c$ from $x$ and $r$
\STATE Encode $c$ into generation-oriented channel symbols $z$
\STATE Transmit $z$ over the wireless channel and obtain $\hat{z}$
\STATE Decode $\hat{z}$ and align it with the conditional embedding space
\STATE Inject the aligned representation into the generative model
\STATE Generate output $\hat{y}$ using the conditioned diffusion or LLM model
\end{algorithmic}
\end{algorithm}

\textbf{Controlled Generation:} Controlled generation ensures that the receiver’s generative model does not produce unconstrained outputs driven solely by its internal priors, but instead creates content that adheres to the information transmitted by the sender. To support this, the transmitter sends not only highly compressed semantic cues but also auxiliary control signals, e.g., conditional prompts, steering codes in latent space, or task-specific constraints—that regulate the generative process. These controls prevent divergence, enforce consistency with domain or physical constraints, and ensure that the generated output faithfully reflects the transmitted information. Controlled generation thus transforms the receiver from a creative generator into a reliable, task-aligned synthesizer.

\begin{figure*}
    \centering
    \includegraphics[width=1\linewidth]{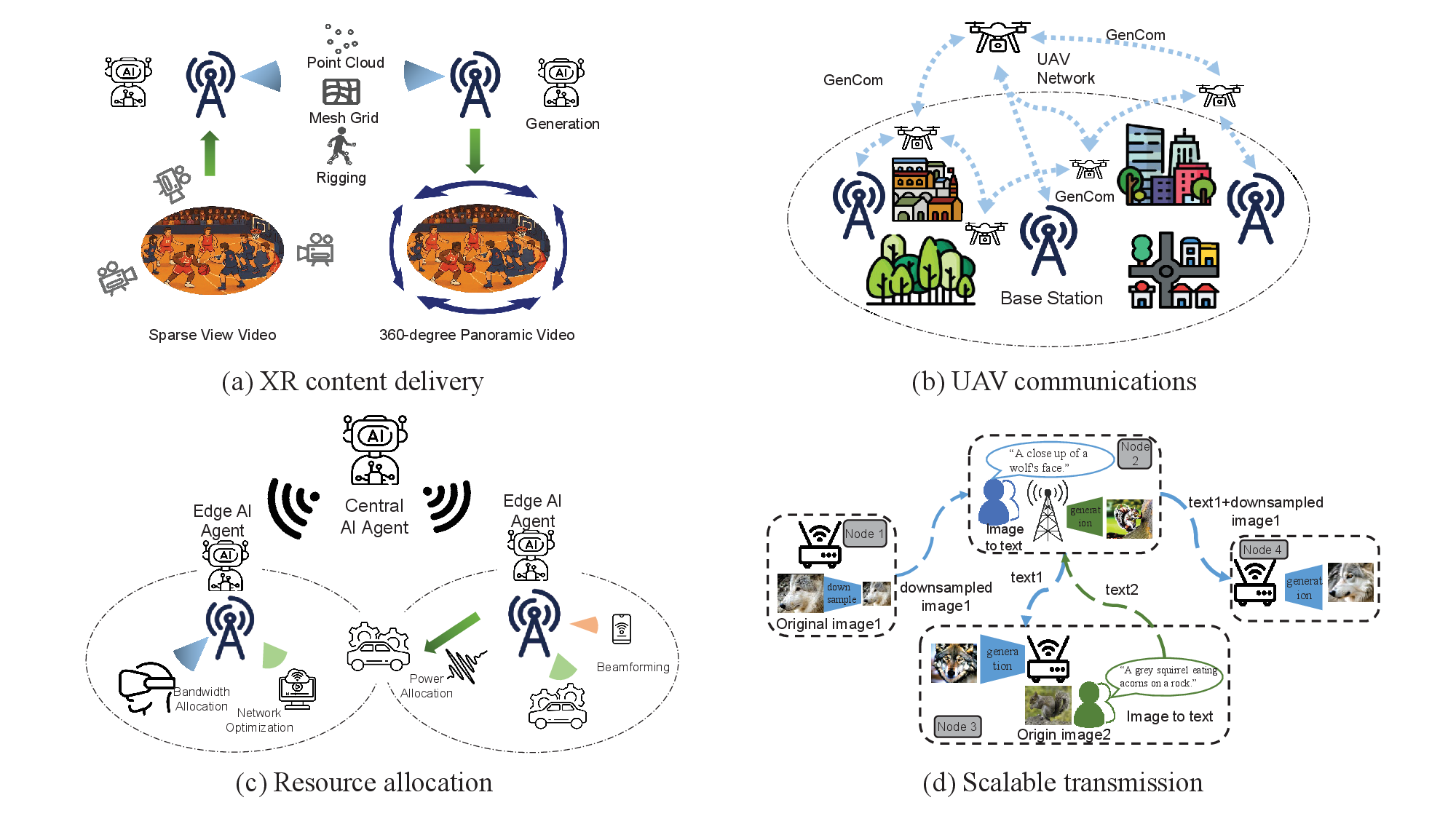}
    \caption{Four typical applications  of GenCom, including XR content delivery, UAV communications, Resource allocation, Scalable transmission.}
    \label{application}
\end{figure*}

\textbf{Communications-aware LLMs:} LLMs serve as the cognitive backbone of GenCom, providing semantic understanding, reasoning, and generation capabilities throughout the communication process. Unlike conventional models, LLMs in GenCom must possess communication awareness, the ability to interpret compact transmission cues, incorporate physical-layer constraints, and dynamically adapt generation according to channel conditions and system requirements. Because 6G devices often have limited computation and memory, these models must also be lightweight, achieved through compression, pruning, quantization, distillation, or edge–device collaborative partitioning. At the control layer, communications-aware LLMs further operate as intelligent agents, performing autonomous resource orchestration, including model scheduling, computational budgeting, and communication optimization. Collectively, these features enable LLMs to support adaptive, efficient, and robust generative communication under realistic network constraints \cite{Wei2026Optimization}.

\textbf{Knowledge-Grounded Generation and Synchronization:} Accurate and interpretable generation in GenCom requires that transmitters and receivers share consistent and up-to-date knowledge bases. This necessitates dedicated mechanisms for knowledge distribution, synchronization, and version control across devices and network edges. However, modern LLM-based knowledge stores can reach gigabyte- or terabyte-scale, making direct transmission over the data plane infeasible. GenCom therefore relies on lightweight synchronization protocols at the control layer, such as model-delta broadcasting (transmitting only parameter or knowledge differences) and incremental knowledge-graph updates. These designs ensure that all endpoints maintain coherent semantic priors while minimizing network overhead, enabling grounded and factually reliable generative outputs \cite{Jiang2024Semantic}.

\textbf{Evaluation Metrics:} Traditional quality of service (QoS) metrics, e.g., BER, mean squared error (MSE), or peak signal-to-noise ratio (PSNR), are insufficient for GenCom because they fail to capture the semantic correctness, usefulness, or factual validity of generated outputs. GenCom therefore requires intent-driven and task-oriented evaluation metrics that align with its generation-centric nature. These include semantic alignment measures (e.g., CLIP score, ViT score) to assess how well the generated content matches the transmitted cues, as well as perceptual metrics (e.g., learned perceptual image patch similarity (LPIPS), fréchet inception distance (FID)) to quantify realism and distortion from a generative perspective. For task-oriented applications, utility-based indicators—such as classification accuracy, control success rate, or mission-completion performance, serve as the definitive measure of communication quality. Importantly, these metrics are not merely used for post-hoc evaluation; they also act as training objectives and real-time feedback signals, guiding the joint optimization of source encoding, channel transmission, and generative reconstruction toward fulfilling the intended task outcome.

GenCom requires intent-driven evaluation. For content-generation-oriented intents, semantic consistency metrics such as CLIP, ViT score, or DINO score measure whether the generated output follows the transmitted cues, while perceptual realism metrics such as FID and KID characterize visual naturalness. For task-execution-oriented intents, the key indicators are task utility metrics such as decision accuracy, control success rate, mission-completion performance, and latency satisfaction. For safety-sensitive intents, safety constraints should be treated as an independent evaluation dimension, because outputs that violate operational or service constraints must be rejected even if they appear semantically plausible.

In practical systems, these metrics should be combined hierarchically into accept or reject rules. Safety constraints are checked first as hard constraints. If they are satisfied, task utility is prioritized for task-oriented applications, whereas semantic consistency and perceptual realism further evaluate generation quality for content-oriented applications.

\subsection{GenCom Versus Related Paradigms}
The boundary between GenCom and semantic communications with generative models. is determined by the primary role of the generative model. If the model is mainly introduced to improve semantic understanding, semantic recovery, or task execution, the scheme remains closer to semantic communication. By contrast, if the model is mainly used for controlled generation of outputs that are better aligned with user or task requirements, or even entirely new rather than directly recovered from the source, the scheme should be understood as GenCom.

This distinction also leads to different transmitted content, coding objectives, receiver functions, and evaluation criteria. Semantic communication mainly transmits task-relevant semantics extracted from existing content and evaluates semantic fidelity or task effectiveness. GenCom instead transmits control or conditioning signals that are sufficient to guide generation, and it evaluates controllability, generation quality, and task outcome jointly. In addition, simply attaching a generator after a conventional joint source-channel coding pipeline at inference time should not be regarded as a native JSCGC, because the coding process itself is not optimized for controlled generation.

GenCom is also different from ``less data, more knowledge'' or knowledge-assisted semantic communication approaches \cite{Chaccour2025Less}. These approaches mainly exploit shared knowledge to reduce the amount of transmitted data and support more efficient inference or semantic reconstruction. GenCom shares this knowledge-driven spirit, but further treats knowledge and generative priors as control interfaces for producing the desired output. Therefore, the design objective is not only payload reduction, but also generation compatibility, controllability, and factual grounding.

Compared with LLM-empowered communication and LLM-native networking frameworks \cite{Jiang2024Large,Wei2026Optimization,Long2025Comprehensive,Jiang2026Agentic,Jiang2026Comprehensive}, GenCom differs in the role assigned to large models. In many LLM-empowered frameworks, LLMs mainly act as intelligent agents for network orchestration, resource allocation, protocol adaptation, or decision-making. In GenCom, large generative models are not only auxiliary tools for network control, but are embedded into the communication process itself: transmitted cues are designed to condition the receiver-side generator, and successful communication is judged by the generated result. Thus, GenCom redefines communication as a generation-oriented process rather than merely enhancing existing communication systems with LLM intelligence.

\section{Typical Applications for GenCom }
GenCom can be applied across a broad range of domains. In this section, we describe four representative application scenarios to illustrate its potential, as shown in Fig. \ref{application}.
\subsection{Extended Reality (XR) Content Delivery}
In XR applications, transmitting full-resolution multi-view or panoramic video is prohibitively expensive due to the massive data rate and stringent latency constraints. In GenCom, instead of sending dense frames, the transmitter extracts compact 3D scene descriptors, such as point clouds, mesh grids, actor rigs, and motion parameters, and communicates only this structured representation. On the receiver, generative models reconstruct high-fidelity 360-degree XR content that accurately reflects the original scene while dramatically reducing communication overhead.

As illustrated in Fig. \ref{application}(a), sparse-view video captured from a limited number of cameras is first encoded into a lightweight representation at the network edge. The receiver then leverages generative reconstruction to synthesize a full panoramic XR experience, ensuring spatial consistency and immersive realism. This representation–generation pipeline enables scalable, low-latency XR delivery for applications such as live sports broadcasting, and interactive entertainment \cite{Ning2025When}.

\subsection{Collaborative UAV Communications}
In multi-UAV systems, transmitting raw sensory data (e.g., images, LiDAR, or video) among drones or toward ground stations is often impractical due to limited airborne power, dynamic channels, and intermittent connectivity. GenCom enables UAVs to exchange only compact semantic intent, such as detected object categories, risk regions, trajectory intent, or mission state codes, while generative models on partnering UAVs or at the edge reconstruct full contextual information.

As shown in Fig. \ref{application}(b), a UAV needs to transmit only the target and its location, while the base station reconstructs the scene by generating the target at the corresponding position based on a real-world map. Under this control, the generated view remains consistent with the physical environment and avoids hallucinations that could mislead mission-critical decision-making. Consequently, GenCom provides a scalable communication framework for large-scale UAV swarms operating in fast-changing real-time environments \cite{Sun2025UAV}.

\subsection{Agent-Based Resource Allocation of Base Station}

In GenCom, base stations can be equipped with LLM-based autonomous agents capable of interpreting compact network descriptors, such as user intent, generative workload type, channel state information, and resource availability, and making real-time resource allocation decisions. These agents understand communication constraints and generative requirements simultaneously, enabling the joint optimization of spectrum assignment, beamforming, model placement, and computational scheduling.

For example, as illustrated in Fig. \ref{application}(c), the edge agent in the left cell dynamically generates bandwidth allocation strategies to support high-fidelity VR transmission while executing network optimization tasks. At the same time, the edge agent in the right cell produces precise power allocation plans for vehicular links and constructs beamforming vectors for mobile users. This collaborative architecture replaces rigid rule-based protocols with generative intelligence, allowing the network to autonomously adapt to diverse service intents and highly dynamic channel conditions \cite{Bariah2024Big}.
\begin{figure}
    \centering
    \includegraphics[width=1\linewidth]{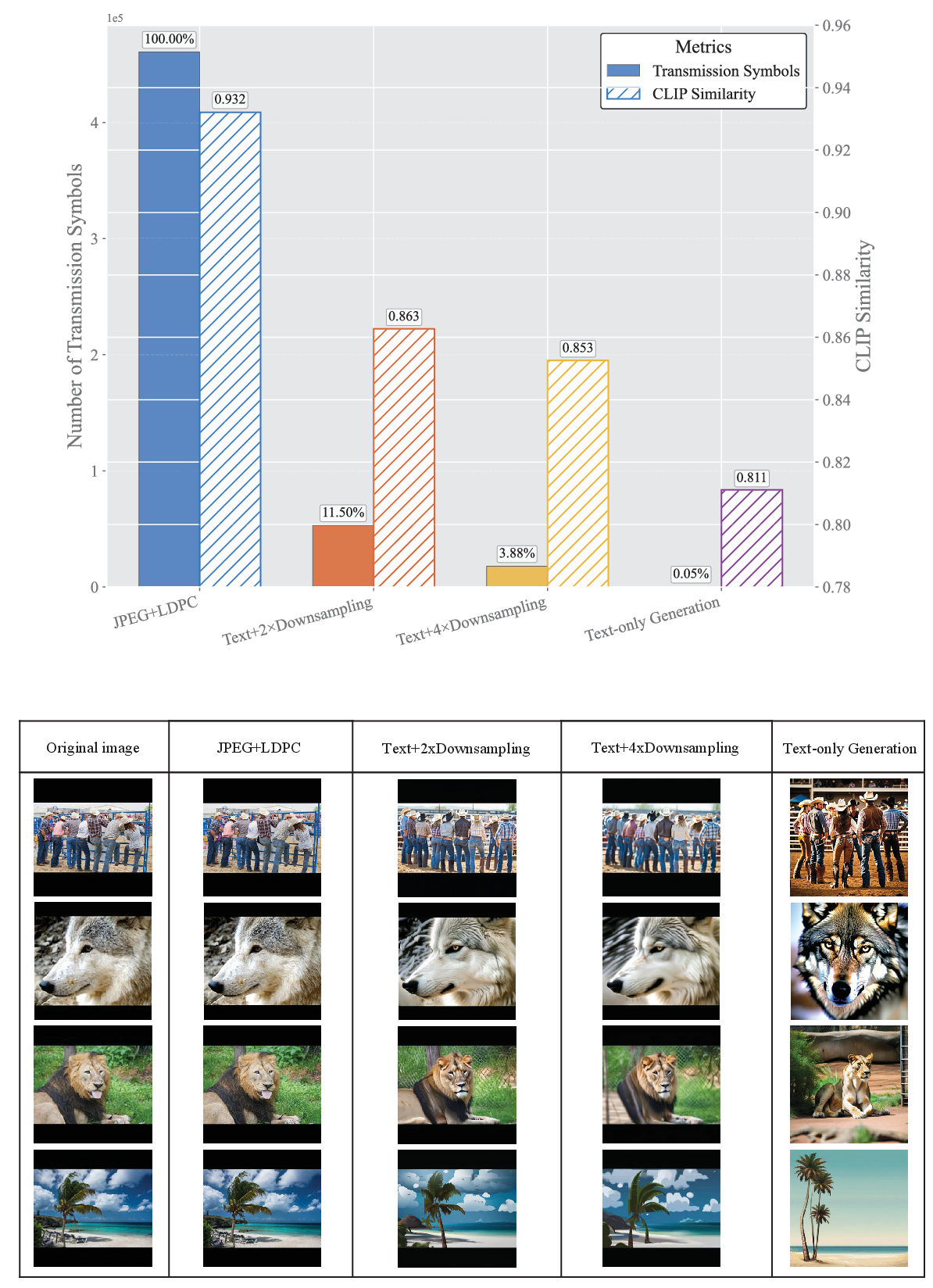}
    \caption{Performance of GenCom under different transmission strategies.}
    \label{fig:result}
\end{figure}

\subsection{Adaptive Semantic-Fidelity Scalable Transmission}

GenCom enables transmission nodes to dynamically adjust the granularity of transmitted content based on network bandwidth, user requirements, and the computational capability of devices. Instead of rigidly sending full-resolution data, the transmitter can scale between high-level semantics and complete content representations. When bandwidth is limited or downstream devices have restricted processing power, only essential information, such as semantic descriptions, scene outlines, or partially downsampled observations—is transmitted. The receiving node then employs LLMs to reconstruct the required content with sufficient fidelity. Conversely, when network resources are abundant, the transmitter can deliver full-resolution data directly, bypassing generative reconstruction and achieving lossless, high-fidelity delivery.

As illustrated in Fig. \ref{application}(d), Node 1 transmits a downsampled image to Node 2, which converts it into a concise semantic description. Node 3 generates a corresponding image from the received semantics and simultaneously transmits its own semantic summary of another scene. These semantic messages, combined with downsampled image, are forwarded to Node 4, which uses generative models to reconstruct a high-quality version of the original content. The figure demonstrates how different nodes flexibly switch between transmitting partial visual information, semantic descriptions, or complete media data depending on channel conditions . 

We conduct experiment of the scalable transmission and report the results in Fig. \ref{fig:result}. The results are obtained on the DIV2K validation set, which contains 200 images. The generative model is SDXL, and the transmitter adopts two representative control formats depending on the transmission requirement. For the text-conditioned case, the input image is first converted into a textual description by BLIP and then encoded for transmission. For the image-conditioned case, the input image is downsampled into a low-resolution image, and JPEG is used as the low-resolution representation to ensure a fair comparison with the baseline. We consider an effectively error-free digital link enabled by channel coding and retransmission, such as LDPC with retransmission, so the comparison focuses on transmission payload rather than SNR-dependent distortion. At the receiver, channel decoding and source decoding recover the transmitted text or low-resolution image, and the recovered control signal is then injected into SDXL through its native conditioning interface. The reported CLIP similarity is computed as the cosine similarity between the CLIP embeddings of the generated image and the reference image, averaged over the whole test set. Since this figure is intended as an overview-level proof of concept, we report average results without additional error bars.

Fig. \ref{fig:result} verifies that GenCom can preserve task-relevant semantics with extremely low transmission cost. The upper plot compares the transmission symbol overhead and CLIP similarity for different schemes, with traditional JPEG+LDPC scheme used as the baseline. The baseline exhibits the highest transmission cost, while Text-only Generation scheme reduces the payload to about 0.05\% and still reaches a CLIP score of 0.811. The Text+Downsampling schemes achieve a similarity range of 0.853 to 0.863 while requiring only 3\% to 12\% of the transmission cost, which shows a clear tradeoff between transmission overhead and semantic fidelity. The lower plot presents the visualization results. We can observe that the JPEG+LDPC scheme provides pixel-level recovery, while Text+Downsampling schemes maintain strong consistency in both content and structure. However, Text-only Generation scheme remains semantically correct but introduces a noticeable style shift. This highlights the importance of controlled generation in GenCom.

\section{Future Research Directions}
\subsection{Towards a Generative Foundational Theory}
Although GenCom has demonstrated substantial gains in bandwidth efficiency and scalability, its theoretical foundations remain largely unexplored. Traditional information-theoretic frameworks, which quantify signal fidelity, are insufficient for generative communications, where the goal is not mere accuracy but the production of outputs that satisfy specific generative objectives. Establishing a rigorous theory requires formalizing the notion of \emph{generative information} as the minimal information necessary to achieve a desired generative outcome, and extending classical concepts such as mutual information and capacity to account for stochasticity, structured dependencies, and task-specific constraints.

A foundational theory would also characterize the \emph{communication-generation capacity}, defining the limits of achievable generative performance under constrained transmission budgets. It would provide principled insights into tradeoffs among transmission overhead, controllability, and task performance, guiding the design of efficient coding, transmission strategies, and evaluation metrics. By bridging information theory, machine learning, and communication systems, such a framework would not only quantify fundamental limits but also inform the development of practical GenCom architectures that optimize efficiency while preserving generative fidelity and task effectiveness.

\subsection{Real-time Processing}
While GenCom significantly alleviates bandwidth pressure through extreme compression, it introduces a new bottleneck in the form of computational inference latency. The current separation of communication and computation optimization is ill-suited for this paradigm, as the time saved in transmission may be negated by the heavy processing required for LLMs at the receiver. Addressing this challenge requires a co-design of the inference-transmission strategy, treating transmission time and generation time as a unified resource budget. Research directions include developing lightweight, few-shot generative models (e.g., consistency models) tailored for devices and designing adaptive split-inference protocols that dynamically distribute generative workloads based on real-time channel conditions and computational availability. Such a unified approach is essential to unlocking the potential of GenCom for latency-sensitive, real-time applications.

\subsection{Collaborative Edge Inference}
In 6G networks, dense and dynamic scenarios such as UAV swarms or vehicular systems require decentralized intelligence, where distributed LLMs collaboratively share intermediate representations, including latent vectors or shared states, to accomplish tasks beyond any single node’s computational or observational capacity. GenCom enables this by transmitting only the generative information necessary for task execution, reducing communication overhead while maintaining high-quality collaborative outputs.

Enabling collaborative edge inference requires communication protocols and coordination strategies for LLM agents. These models must jointly interpret global network states, optimize resources, and perform seamless task handovers while balancing communication overhead, latency, and collective generative performance. By allowing distributed LLMs to reason and communicate adaptively, decentralized GenCom can achieve scalable, resilient, and efficient generative intelligence across edge networks \cite{WDMoE}.

\subsection{Security and Semantic Resilience}
GenCom introduces new attack surfaces within the high-dimensional latent space, making traditional security measures insufficient. Adversarial attacks targeting transmitted semantic information or the generative model itself could inject imperceptible perturbations, causing the receiver to synthesize erroneous or malicious outputs. Such threats pose critical risks to system integrity and safety.

Future research must prioritize the development of semantic resilience methods. This includes defenses that detect and neutralize adversarial perturbations in latent embeddings before they impact generation. Moreover, theoretical approaches are needed to prevent privacy leakage to ensure that the semantic representations are protected from unauthorized inference. Developing robust semantic encryption, authentication, and access-control protocols is essential for GenCom.

\section{Conclusion}
In this article, we introduced GenCom as a new communication paradigm that shifts the design objective from ``transmitting everything accurately” to ``transmitting only what is necessary for controlled generation.” We presented the definition of GenCom as embedding understanding, reasoning, and generation directly into the communication process, enabling the transmitter and the receiver to collaboratively produce task-relevant outcomes. A two-layer architecture was designed for GenCom, which is supported by key enabling technologies such as joint source–channel–generative coding, controlled generation, communications-aware LLMs, knowledge synchronization, and evaluation metrics. Four typical application scenarios, including XR transmission and UAV communications, were provided to illustrate the potential of GenCom in 6G networks. We also discussed future research directions for GenCom, addressing both fundamental theories and practical areas such as real-time processing.

\section{Acknowledgments}
This work is supported by the National Natural Science Foundation of China (NSFC) under Grant 62431015, and Grant 625B2112, and Grant 62125108. 
\bibliographystyle{IEEEtran}
\bibliography{ref}

@article{clark2013whatever,
  title={{Whatever next? Predictive brains, situated agents, and the future of cognitive science}},
  author={Clark, A},
  journal={Behavioral and Brain Sciences},
  year={2013},
  volume={36},
  pages={181--204}
}

@ARTICLE{Jiang2024Large,
  author={Jiang, Feibo and Peng, Yubo and Dong, Li and Wang, Kezhi and Yang, Kun and Pan, Cunhua and Niyato, Dusit and Dobre, Octavia A.},
  journal={IEEE Wireless Communications}, 
  title={{Large Language Model Enhanced Multi-Agent Systems for 6G Communications}}, 
  year={2024},
  volume={31},
  number={6},
  pages={48-55},
  doi={10.1109/MWC.016.2300600}}

@article{ZHANG2025Generative,
title = {{Generative Video Communications: Concepts, Key Technologies, and Future Research Trends}},
journal = {Engineering},
year = {2025},
issn = {2095-8099},
doi = {https://doi.org/10.1016/j.eng.2025.06.018},
author = {Wenjun Zhang and Guo Lu and Zhiyong Chen and Geoffrey Ye Li},
}

@ARTICLE{Chaccour2025Less,
  author={Chaccour, Christina and Saad, Walid and Debbah, Mérouane and Han, Zhu and Vincent Poor, H.},
  journal={IEEE Communications Surveys \& Tutorials}, 
  title={{Less Data, More Knowledge: Building Next-Generation Semantic Communication Networks}}, 
  year={2025},
  volume={27},
  number={1},
  pages={37-76},
  doi={10.1109/COMST.2024.3412852}}

@ARTICLE{Bariah2024Big,
  author={Bariah, Lina and Zhao, Qiyang and Zou, Hang and Tian, Yu and Bader, Faouzi and Debbah, Merouane},
  journal={IEEE Communications Magazine}, 
  title={{Large Generative AI Models for Telecom: The Next Big Thing?}}, 
  year={2024},
  volume={62},
  number={11},
  pages={84-90},
  doi={10.1109/MCOM.001.2300364}}

@ARTICLE{Jiang2024Semantic,
  author={Jiang, Feibo and Peng, Yubo and Dong, Li and Wang, Kezhi and Yang, Kun and Pan, Cunhua and You, Xiaohu},
  journal={IEEE Wireless Communications}, 
  title={{Large AI Model-Based Semantic Communications}}, 
  year={2024},
  volume={31},
  number={3},
  pages={68-75},
}

@ARTICLE{CDDM,
  author={Wu, Tong and Chen, Zhiyong and He, Dazhi and Qian, Liang and Xu, Yin and Tao, Meixia and Zhang, Wenjun},
  journal={IEEE Transactions on Wireless Communications}, 
  title={{CDDM: Channel Denoising Diffusion Models for Wireless Semantic Communications}}, 
  year={2024},
  volume={23},
  number={9},
  pages={11168-11183},
}

@ARTICLE{WDMoE,
  author={Xue, Nan and Sun, Yaping and Chen, Zhiyong and Tao, Meixia and Xu, Xiaodong and Qian, Liang and Cui, Shuguang and Zhang, Wenjun and Zhang, Ping},
  journal={IEEE Transactions on Wireless Communications}, 
  title={{WDMoE: Wireless Distributed Mixture of Experts for Large Language Models}}, 
  year={2025},
  volume={},
  number={},
  pages={1-1},
}

@article{Ning2025When,
author = {Xinyu Ning and Yan Zhuo and Xian Wang and Chan-In Devin Sio and Lik-Hang Lee},
title = {{When Generative Artificial Intelligence Meets Extended Reality: A Systematic Review}},
journal = {International Journal of Human–Computer Interaction},
volume = {0},
number = {0},
pages = {1--21},
year = {2025},
publisher = {Taylor \& Francis},
}

@ARTICLE{Sun2025UAV,
  author={Sun, Geng and Xie, Wenwen and Niyato, Dusit and Du, Hongyang and Kang, Jiawen and Wu, Jing and Sun, Sumei and Zhang, Ping},
  journal={IEEE Network}, 
  title={{Generative AI for Advanced UAV Networking}}, 
  year={2025},
  volume={39},
  number={4},
  pages={244-253},
}

@ARTICLE{Long2025Comprehensive,
  author={Long, Shuowen and others},
  journal={IEEE Network},
  title={{6G Comprehensive Intelligence: Network Operations and Optimization Based on Large Language Models}},
  year={2025},
  volume={39},
  number={4},
  pages={192-201},
}

@ARTICLE{Wei2026Optimization,
  author={Wei, Bowen and others},
  journal={IEEE Communications Surveys \& Tutorials},
  title={{Large Language Models for Optimization in Next-Generation Wireless Network Management: A Survey}},
  year={2026},
  volume={28},
  pages={5713-5746},
}

@ARTICLE{Jiang2026Agentic,
  author={Jiang, Feibo and Pan, Cunhua and Wang, Kezhi and Michiardi, Pietro and Dobre, Octavia A. and Debbah, Merouane},
  journal={IEEE Journal on Selected Areas in Communications},
  title={{From Large AI Models to Agentic AI: A Tutorial on Future Intelligent Communications}},
  year={2026},
  volume={44},
  pages={3507-3540},
}

@ARTICLE{Jiang2026Comprehensive,
  author={Jiang, Feibo and others},
  journal={IEEE Communications Surveys \& Tutorials},
  title={{A Comprehensive Survey of Large AI Models for Future Communications: Foundations, Applications, and Challenges}},
  year={2026},
  volume={28},
  pages={4731-4764},
}

@ARTICLE{JSCGC,
  author={Wu, Tong and Chen, Zhiyong and Lu, Guo and Song, Li and Yang, Feng and Tao, Meixia and Zhang, Wenjun},
  journal={2026 IEEE International Symposium on Information Theory Workshops}, 
  title={{Joint Source-Channel-Generation Coding: From Distortion-oriented Reconstruction to Semantic-consistent Generation}},
  year={2026},

}

\section*{Biographies}
\begin{IEEEbiographynophoto}{Wenjun Zhang}
   received his B.S., M.S. and Ph.D. degrees in electronic engineering from Shanghai Jiao Tong University, Shanghai, China, in 1984, 1987 and 1989, respectively. After three years’ working as an engineer at Philips in Nuremberg, Germany, he went back to his Alma Mater in 1993 and became a full professor of Electronic Engineering in 1995. He was one of the main contributors of the Chinese DTTB Standard (DTMB) issued in 2006. He holds 238 patents and published more than 130 papers in international journals and conferences. He is the Chief Scientist of the Chinese Digital TV Engineering Research Centre (NERC-DTV), an industry/government consortium in DTV technology research and standardization, and the director of Cooperative Media Network Innovation Centre (CMIC), an excellence research cluster affirmed by the Chinese Government. He is an Academician of the Chinese Academy of Engineering (CAE).  His main research interests include video coding and wireless transmission, multimedia semantic analysis and broadcast/broadband network convergence. 
\end{IEEEbiographynophoto}
\vspace{-1cm}

\begin{IEEEbiographynophoto}{Zhiyong Chen}
  received the Ph.D. degree from the School of Information and Communication Engineering, Beijing University of Posts and Telecommunications (BUPT), Beijing, China, in 2011. From 2009 to 2011, he was a visiting Ph.D. Student at the Department of Electronic Engineering, University of Washington, Seattle, USA. He is currently a Professor with the Cooperative Medianet Innovation Center, Shanghai Jiao Tong University (SJTU), Shanghai, China. His research interests include mobile communications-computing-caching (3C) networks and mobile AI systems. He served as the Student Volunteer Chair for the IEEE ICC 2019, the Publicity Chair for the IEEE/CIC ICCC 2014 and a TPC member for major international conferences. He was the recipient of the IEEE Asia-Pacific Outstanding Paper Award in 2019.
\end{IEEEbiographynophoto}
\vspace{-1cm}

\begin{IEEEbiographynophoto}{Tong Wu}
  received the B.S degree in Electronic and Information Engineering from Dalian University of Technology, Dalian, China, in 2022. He is currently pursuing the Ph.D degree with the Department of Information and Communication Engineering. His research interests include generative semantic communications and machine learning for wireless networks.
\end{IEEEbiographynophoto}
\vspace{-1cm}

\begin{IEEEbiographynophoto}{Guo Lu}
(Member, IEEE) received his B.S. degree in Electronic Engineering from Ocean University of China, China, in 2014, and his Ph.D. degree in Electronic Engineering from Shanghai Jiao Tong University (SJTU), China, in 2020. He is currently an Associate Professor at SJTU. His research interests focus on learned video coding and processing. Dr. Lu has published over 40 papers in prestigious journals and conferences, including T-IP, T-CSVT, CVPR, and T-PAMI. He is a recipient of the 2023 IEEE CASS Visual Signal Processing and Communications (VSPC) Rising Star Award, the China Society of Image and Graphics (CSIG) Excellent Doctoral Dissertation Award (Top 10 Nationwide), and the SJTU Excellent Doctoral Dissertation Award. He has served as a Guest Editor for IJCV and IEEE T-CSVT, is a member of the IEEE VSPC Technical Committee, and has organized several tutorials on learned video compression at CVPR, ACMMM, and VCIP. Additionally, he served as Publication Chair for MLSP and as a Senior Program Committee member for AAAI.
\end{IEEEbiographynophoto}
\vspace{-1cm}

\begin{IEEEbiographynophoto}{Li Song}
    (Senior Member, IEEE) received the B.E. and M.S. degrees in engineering in 1997 and 2000, respectively, and the Ph.D. degree in electrical engineering from Shanghai Jiao Tong University (SJTU) in 2005. He is currently a full professor with the department of electronic engineering. He was also a visiting professor with Santa Clara University from 2011 to 2012. He has 300 publications, 50 granted patents, and 20 standard technical contributions. His research interests include visual signal processing and artificial intelligence for multimedia. He was a recipient of the National Science and Technology Progress Award in 2015, the Okawa Foundation Research Grant in 2012, the Best Paper Award of the IEEE Circuits and Systems Society Visual Signal Processing and Technical Committee (IEEE CASS VSPCTC) in 2022, the Second Place Award from IEEE ICME-Twitch Grand Challenge in 2017, the Best 10\% Paper Award from the IEEE VCIP in 2016, and the Best Paper Award from the WCSP in 2010. He has been serving as an associate editor for Multidimensional Systems and Signal Processing from 2012 to 2018 and an associate editor for the IEEE Transactions on Broadcasting since 2024.
\end{IEEEbiographynophoto}
\vspace{-1cm}

\begin{IEEEbiographynophoto}{Feng Yang}
  (Member, IEEE) received the Ph.D. degree in information and communication from Shanghai Jiao Tong University. Since 2008, he has been a Faculty Member with Shanghai Jiao Tong University, where he is currently a Professor with the Department of Electronic Engineering. His research interests include wireless communication and array signal processing.
\end{IEEEbiographynophoto}
\vspace{-1cm}

\begin{IEEEbiographynophoto}{Meixia Tao}
  (F'19) is a Distinguished Professor with the School of Information Science and Electronic Engineering at Shanghai Jiao Tong University, China. She received the B.S. degree in electronic engineering from Fudan University, Shanghai, China, in 1999, and the Ph.D. degree in electrical and electronic engineering from Hong Kong University of Science and Technology in 2003. Her current research interests include wireless edge learning, semantic communications, integrated communication-computing-sensing, AI-based channel modeling and beamforming. Dr. Tao receives the 2019 IEEE Marconi Prize Paper Award, the 2013 IEEE Heinrich Hertz Award for Best Communications Letters, the IEEE/CIC International Conference on Communications in China (ICCC) 2015 Best Paper Award, and the International Conference on Wireless Communications and Signal Processing (WCSP) 2022 and 2012 Best Paper Awards. She also receives the 2009 IEEE ComSoc Asia-Pacific Outstanding Young Researcher award.
\end{IEEEbiographynophoto}
\vspace{-1cm}

\end{document}